\documentclass[twocolumn,showpacs,prb]{revtex4-1}%
\usepackage{amsfonts}
\usepackage{amsmath}
\usepackage{amssymb}
\usepackage{subfigure}
\usepackage{graphicx}
\usepackage{txfonts}%
\providecommand{\U}[1]{\protect\rule{.1in}{.1in}}

\begin{document}

\title{Veselago focusing of anisotropic massless Dirac fermions}
\author{Shu-Hui Zhang$^{1}$}
\email{shuhuizhang@mail.buct.edu.cn}
\author{Wen Yang$^{2}$}
\email{wenyang@csrc.ac.cn}
\author{F. M. Peeters$^{3}$}
\email{francois.peeters@uantwerpen.be}

\affiliation{$^{1}$College of Science, Beijing University of Chemical Technology, Beijing,
100029, China}
\affiliation{$^{2}$Beijing Computational Science Research Center, Beijing 100193, China}
\affiliation{$^{3}$Department of Physics, University of Antwerp, Groenenborgerlaan 171, B-2020 Antwerpen, Belgium}

\begin{abstract}

Massless Dirac fermions (MDFs) emerge as quasiparticles in various novel materials such as graphene and topological insulators, and they exhibit several intriguing properties, of which Veselago focusing is an outstanding example with a lot of possible applications. However, up to now Veselago focusing merely occurred in \textit{p-n} junction devices based on the isotropic MDF, which lacks the tunability needed for realistic applications. Here, motivated by the emergence of novel Dirac materials, we investigate the propagation behaviors of anisotropic MDFs in such a \textit{p-n} junction structure. By projecting the Hamiltonian of the anisotropic MDF to that of the isotropic MDF and deriving an exact analytical expression for the propagator, precise Veselago focusing is demonstrated without the need for mirror symmetry of the electron source and its focusing image. We show a tunable focusing position that can be used in a device to probe masked atom-scale defects. This study provides an innovative concept to realize Veselago focusing relevant for potential applications, and it paves the way for the design of novel electron optics devices by exploiting the anisotropic MDF.

\end{abstract}
\maketitle

\section{Introduction}

Massless Dirac fermions (MDFs) have emerged as quasiparticles in many novel materials, such as graphene \cite{CastroRMP2009} and topological insulators \cite{QiRMP2011}. The unique physics and fascinating phenomena of MDFs motivate the search for new Dirac materials \cite{Wehling2014,WangNSRl2015}, which usually have novel energy dispersions. For example, anisotropic MDFs exist extensively in two-dimensional \cite{Kobayashi2007,PhysRevLett.103.016402,PhysRevLett.104.137001,PhysRevLett.107.086801,PhysRevLett.113.156602,PhysRevB.84.195425,PhysRevB.93.241405,li2017}
and three-dimensional Dirac materials
\cite{PhysRevLett.107.126402,PhysRevLett.115.026403,Yan2017}. In addition, the MDFs in graphene can also be tuned from isotropic to anisotropic by strain \cite{PereiraPRL2009,Naumis2017}, the application of a superlattice potential \cite{Park2008,PhysRevB.81.075438,PhysRevLett.101.126804,PhysRevLett.105.246803}, and by partial hydrogenation \cite{PhysRevB.94.195423}.
There is continuing enthusiasm to exploit the unusual transport properties of MDFs in various host systems, which are crucial
for future potential applications \cite{Wehling2014}.

Due to MDFs' unique features of being gapless and high mobility, they are ideal to realize different electron optics applications \cite{CheianovScience2007,WilliamsNatNano2011,RickhausNatComm2013,ncomms7093}, of which Veselago focusing is an outstanding example \cite{CheianovScience2007}. This seminal study \cite{CheianovScience2007} brings the concept of negative refraction into graphene and then provides an electronic analog of Veselago focusing. This remarkable theoretical result stood as a challenge to experimentalists \cite{Pendry2007}. Veselago focusing implies that all electron waves diverging from a source across the junction converge into a focal image due to negative refraction, which lies at the heart of many theoretical proposals  \cite{CheianovScience2007,PhysRevLett.100.236801,MoghaddamPRL2010,SilveirinhaPRL2013,ZhaoPRL2013,MilovanovicJAP2015,ncomms15783,zhang2017,PhysRevB.95.214103}. In particular, Veselago focusing has been observed in two recent experiments \cite{LeeNatPhys2015,ChenScience2016}, which
boosted new research interest.

Previous studies were either limited to the isotropic MDF \cite{CheianovScience2007,ZhaoPRL2013,PhysRevB.95.214103} or they showed that the anisotropy of energy dispersion deteriorates Veselago focusing \cite{HasslerPRB2010,intraband2012}. Therefore, it seems impossible to exceed  the requirement of the isotropic MDF, which leads to limited electronic systems realizing Veselago focusing, and a lack of tunability for relevant applications. For example, the source is usually fixed, which leads to an immovable focal image at its mirror position \cite{CheianovScience2007}. In this study, we consider Veselago focusing of anisotropic MDFs using a \textit{p-n} junction (PNJ) structure. By projecting the Hamiltonian of the anisotropic MDF to that of the isotropic MDF, we derive an exact analytical expression for the propagator in order to show the precise Veselago focusing, which we show to have superior tunable features. The tunable features not only lead to a novel design, e.g., to probe the masked defect by utilizing a tunable focusing position, but they also favor the previous proposed applications based on Veselago focusing. This study presents an innovative concept to realize Veselago focusing that will be beneficial for potential applications, and it provides another way to design electron optics devices by utilizing anisotropic MDFs.

\begin{figure}[htbp]
\includegraphics[width=\columnwidth,clip]{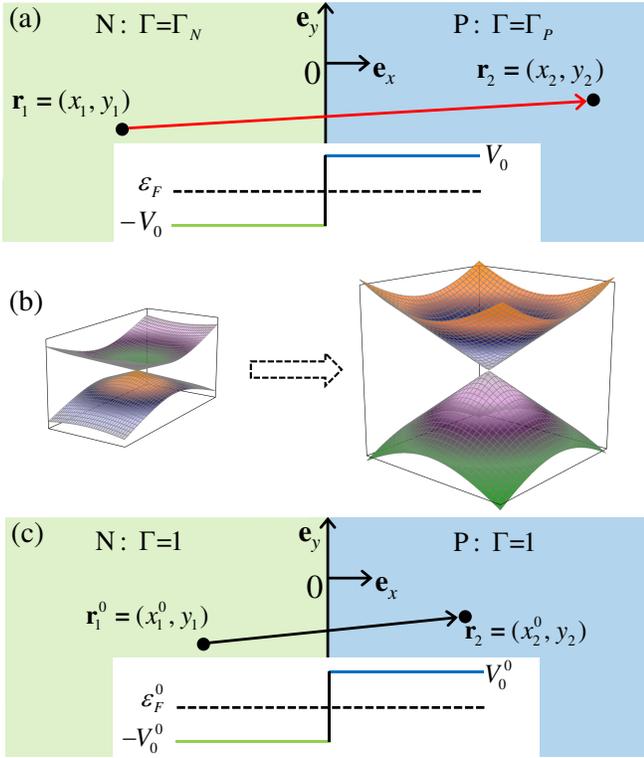}\caption{The
\textit{p-n} junctions based on the anisotropic and isotropic massless Dirac
fermions, which can be projected into each other. (a) The \textit{p-n} junction
based on anisotropic massless Dirac fermions, in which N and P regions
have anisotropy $\Gamma=\Gamma_{N}$ and $\Gamma=\Gamma_{P}$,
respectively. To investigate the propagation properties, we consider the Green
function or propagator denoted by the red line with an arrow from
$\mathbf{{r}}_{1}=(x_{1},y_{1})$ to $\mathbf{{r}}_{2}=(x_{2},y_{2})$ in the
Cartesian coordinate system ($\mathbf{{e}}_{x}$, $\mathbf{{e}}_{y}$). The
inset in the white rectangle zone shows the Fermi level $E_{F}$ relative to
the Dirac points with the energy positions $-V_{0}$ in the N region and $V_{0}$ in
the P region. (b) The anisotropic massless Dirac
fermions can be projected into the isotropic ones. (c) The \textit{p-n} junction
based on the isotropic massless Dirac fermions with $\Gamma=1$, which is
projected from (a). Here, we used $\Gamma_{N}=5/8$ and
$\Gamma_{P}=1/2$.}%
\label{structure}%
\end{figure}

\section{Theoretical formalism}

\subsection{Model and Hamiltonian}

The considered PNJ structure is shown
schematically in Fig. \ref{structure}(a) and it consists of a left N region and
a right P region. Each region of the PNJ hosts the anisotropic MDF for which the
anisotropy can be different in the N and P regions. In general, the
Hamiltonian of the PNJ in Fig. \ref{structure}(a) has the form:%

\begin{equation}
\hat{H}=(\hat{H}_{N}+V_{N})\Theta(-x)+(\hat{H}_{P}+V_{P})\Theta(x),
\end{equation}
where $\hat{H}_{i}$ is the intrinsic Hamiltonian of the $i$ ($=$P, N) region, $V_{N}=-V_{0}$ ($V_{P}=V_{0}$) is the gate-induced scalar
potential in the N (P) region by assuming $V_{0}>0$ without loss of generality,
and $\Theta(x)$ is the step function: $\Theta(x)=1$ for $x>0$ and
$\Theta(x)=0$ for $x<0$. In general, the anisotropic MDF of each uniform
region can be described by using the Hamiltonian \cite{Naumis2017} $\hat
{H}_{i}=v_{F}(\hat{\sigma}_{x}\hat{p}_{x}+\Gamma_{i}\hat{\sigma}_{y}\hat
{p}_{y})$. Here, $v_{F}$ represents the Fermi velocity, and $\hat{\sigma
}_{x,y}$ and $\hat{p}_{x,y}$ are, respectively, the Pauli operator and the
momentum operator. In particular, $\Gamma=\Gamma_{i}$ is introduced to account
for the anisotropy of the MDF in the $i$ region, and the specific value of anisotropy depends on the choice of material and the way to tune the energy dispersion, e.g., for the MDF in graphene, the anisotropy can be tuned continuously by reversible strain up to a factor of 5 \cite{PhysRevB.80.045401} and may be even larger by using a superlattice potential \cite{Park2008}. For the intrinsic Hamiltonian, the energy
dispersion is $\varepsilon_{\mu}%
(\mathbf{k}_{i})=\mu v_{F}\sqrt{k_{i,x}^{2}+\Gamma_{i}^{2}k_{y}^{2}}$ where
the index $\mu=+$ ($\mu=-$) is for the conductance (valence) band,
$\mathbf{k}_{i}=(k_{i,x},k_{y})$ is the momentum
vector, and the corresponding position vector is
$\mathbf{r}=(x,y)$. Note that we take units such that $\hbar\equiv1$ throughout this work.

\subsection{Green's function and the projection method}

To investigate the
propagation properties of the anisotropic MDF in the PNJ structure, we
concentrate on the corresponding propagator or Green's function (GF),
which is defined as $\mathbf{G}(\mathbf{r}_{2},\mathbf{r}_{1},\varepsilon
_{F},V_{0})=\langle\mathbf{r}_{2}|(\varepsilon_{F}+i0^{+}-\hat{H}%
)^{-1}|\mathbf{r}_{1}\rangle$ shown by the red line with an arrow in Fig.
\ref{structure}(a). Note that $\mathbf{G}$ is a matrix due to the spinor
nature of $\hat{H}$. We have developed a simple and elegant method to derive
the GF of isotropic MDF in graphene PNJ structure through the matching
technique combining translational invariance along the interface direction
of the junction \cite{zhang2017}. The generalization of this method to the
anisotropic MDF is straightforward. However, here we present an alternative
but more simple method, i.e, the projection method, which can give the PNJ GF
of the anisotropic MDF from that of the isotropic MDF one. To this aim, we project the
anisotropic Hamiltonian $\hat{H}_{i}$ into the form $\hat{H}_{i}^{0}%
=v_{F}(\hat{\sigma}_{x}\hat{p}_{i,x}^{0}+\hat{\sigma}_{y}\hat{p}_{y})$, where
$\hat{H}_{i}^{0}\equiv\hat{H}_{i}/\Gamma_{i}$ is the Hamiltonian for the
isotropic MDF, and $\hat{p}_{i,x}^{0}\equiv
\hat{p}_{x}/\Gamma_{i}$. The corresponding energy dispersion is
$\varepsilon_{\mu}^{0}(\mathbf{k}_{i}^{0})=\mu v_{F}\sqrt{(k_{i,x}^{0}%
)^{2}+k_{y}^{2}}$ where $\varepsilon_{\mu
}^{0}(\mathbf{k}_{i}^{0})=\varepsilon_{\mu}(\mathbf{k}_{i})/\Gamma_{i}$
presents the projection relation for energy dispersion, $\mathbf{k}_{i}%
^{0}\mathbf{=(}k_{i,x}^{0},k_{y}\mathbf{)}$  is the projected momentum vector with
$k_{i,x}^{0}=k_{i,x}/\Gamma_{i}$, and the corresponding position vector is $\mathbf{r}^{0}\mathbf{=(}%
x^{0},y\mathbf{)}$ with $x^{0}=\Gamma_{i}x$. Here,
to obtain the projection relation $x^{0}=\Gamma_{i}x$, we have used the
constraint $[\hat{x},\hat{p}_{x}]=[\hat{x}^{0},\hat{p}_{x}^{0}%
]=i$\ required by performing the Hamiltonian projection. Interestingly, the
projection relation for the energy dispersion changes the MDF from an anisotropic to
an isotropic one as shown by Fig. \ref{structure}(b). As a result, we obtain the
equivalent PNJ of Fig. \ref{structure}(a) but based on the isotropic MDF\ as
shown by\ Fig. \ref{structure}(c)\ through the projection relation for the
position vector in real space, and the projection relation for the energy
dispersion in energy space leads to%

\begin{equation}
\varepsilon_{i}^{0}=\frac{\varepsilon_{i}}{\Gamma_{i}}. \label{SRED}%
\end{equation}

Here, $\varepsilon_{N}^{0}=V_{N}^{0}+\varepsilon_{F}^{0}$ and $\varepsilon
_{P}^{0}=V_{P}^{0}-\varepsilon_{F}^{0}$ ($\varepsilon_{N}=V_{N}+\varepsilon
_{F}$ and $\varepsilon_{P}=V_{P}-\varepsilon_{F}$)\ represent the doping
levels since the Fermi level $\varepsilon_{F}^{0}$ ($\varepsilon_{F}$) lies
between the junction potentials of the N and P regions $\varepsilon_{F}^{0}\in
\lbrack-V_{0}^{0},V_{0}^{0}]$ ($\varepsilon_{F}\in\lbrack-V_{0},V_{0}]$)$\ $in
the PNJ based on the isotropic (anisotropic)\ MDF. Obviously, the doping
level\ defines the momentum through the energy dispersions, e.g.,
$\varepsilon_{i}^{0}(\mathbf{k}_{i}^{0})=v_{F}\sqrt{(k_{i,x}^{0})^{2}%
+k_{y}^{2}}$. In the two equivalent PNJ structures, the GFs of the isotropic and
anisotropic MDFs are related to each other:
\begin{equation}
\mathbf{G}(\mathbf{r}_{2},\mathbf{r}_{1},\varepsilon_{F},V_{0})=\mathbf{G}%
^{0}(\mathbf{r}_{2}^{0},\mathbf{r}_{1}^{0},\varepsilon_{F}^{0},V_{0}^{0}).
\end{equation}

Here, the PNJ GF of the isotropic MDF [see the black line with an arrow in
Fig. \ref{structure} (c)] is defined as $\hat{G}^{0}(\mathbf{r}_{2}%
^{0},\mathbf{r}_{1}^{0},\varepsilon_{F}^{0},V_{0}^{0})\equiv\langle
\mathbf{r}_{2}^{0}|(\varepsilon_{F}^{0}+i0^{+}-\hat{H}^{0})^{-1}%
|\mathbf{r}_{1}^{0}\rangle$ where $\hat{H}^{0}$ is the PNJ Hamiltonian of the
isotropic MDF in the form:%

\begin{equation}
\hat{H}^{0}=(\hat{H}_{N}^{0}+V_{N}^{0}\mathbf{I})\Theta(-x^{0})+(\hat{H}%
_{P}^{0}+V_{P}^{0}\mathbf{I})\Theta(x^{0}).
\end{equation}

\begin{figure}[htp]
\includegraphics[width=0.98\columnwidth,clip]{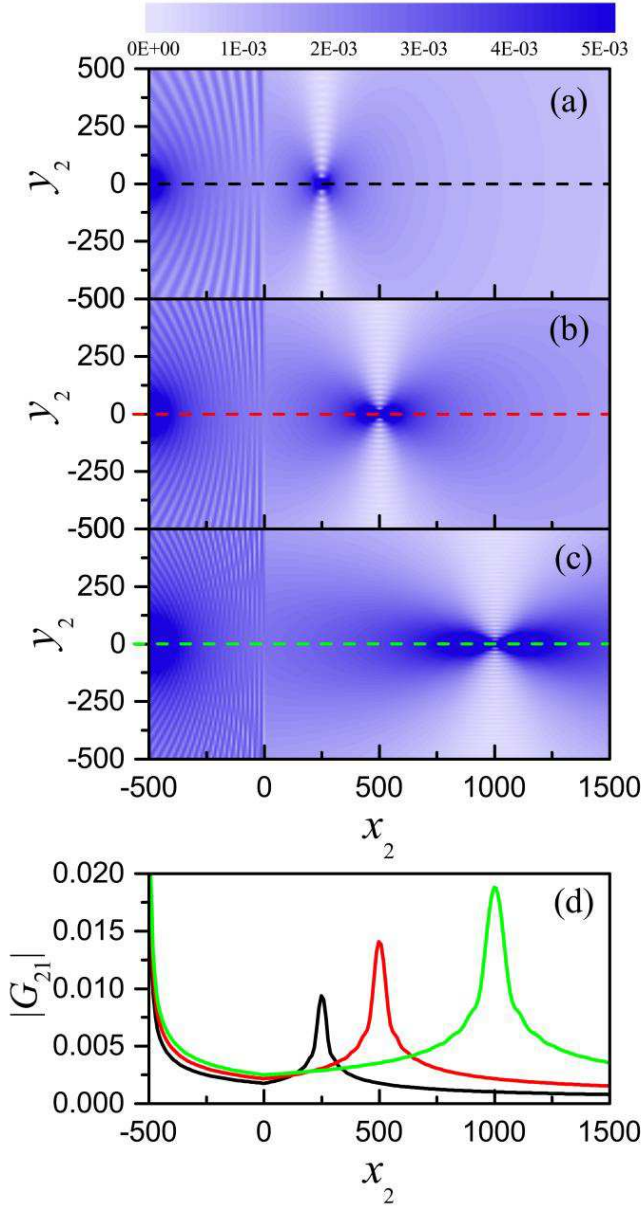}\caption{Veselago
focusing of anisotropic massless Dirac fermions shown by the magnitude of
$|G_{21}|$ as a function of $\mathbf{r}_{2}=(x_{2},y_{2})$. Here, $G_{21}$
is the matrix element of the propagator matrix $\mathbf{G}(\mathbf{r}%
_{2},\mathbf{r}_{1},\varepsilon_{F},V_{0})$. Without loss of generality, we assume
$\mathbf{r}_{1}=(-500,0)$, $V_{0}=0.2$, and $\Gamma_{N}=1$. (a) $\varepsilon
_{F}=-V_{0}/3$, and $\Gamma_{P}=2$. (b) $\varepsilon_{F}=0$, and $\Gamma
_{P}=1$. (c) $\varepsilon_{F}=V_{0}/3$, and $\Gamma_{P}=1/2$. (d) $|G_{21}|$
as a function of $x_{2}$ for $y_{2}=0$ using the same color code as in (a)-(c). In the plot, we define the unit length $a_0$ and unit
energy $t_0$ through $v_{F}=3/2a_0t_0$ by analogy to graphene \cite{CastroRMP2009}.}%
\label{focusing}%
\end{figure}

\subsection{Analytical Green's function of the isotropic MDF}

First, we need
to derive PNJ GF of the isotropic MDF. By examining the propagation phase and its
higher order derivative, we present a detailed analytical derivation of
the PNJ GF of the isotropic MDF in Appendix A, which helps to construct the intuitive physical picture
for the propagation properties of the isotropic MDF across the PNJ, i.e., the
classical trajectories, negative refraction and then Veselago focusing
\cite{CheianovScience2007}. We assume a source at $\mathbf{r}_{1}^{0}=(-a,0)$;
the Veselago focusing occurs at its mirror image $\mathbf{r}_{2}%
^{0}=\mathbf{r}_{1\mathrm{m}}^{0}=(a,0)$ and $\varepsilon_{F}^{0}=0$ in the case of a symmetric junction implying $\varepsilon_{N}^{0}%
=\varepsilon_{P}^{0}$. The analytical formula for the PNJ GF of the isotropic
MDF is $\mathbf{G}^{0}(\mathbf{r}_{1\mathrm{m}}^{0},\mathbf{r}_{1}%
^{0},\varepsilon_{F}^{0},V_{0}^{0})=\mathcal{G}(V_{0}^{0})$ from the definition%

\begin{equation}
\mathcal{G}(V_{0}^{0})\equiv\frac{\rho(V_{0}^{0})}{2i}(\frac{\pi}{2}%
+2\sigma_{x}). \label{IGF}%
\end{equation}
where $\rho(V_{0}^{0})=V_{0}^{0}/(2\pi v_{F}^{2})$ is the density of states of
the isotropic MDF with the doping level $V_{0}^{0}$. Because $\mathcal{G}\propto V_{0}^{0}$ an enhancement of the
focusing intensity of the isotropic MDF occurs when increasing the doping level
through electrical gating or using other ways. On the other hand, the focusing
position has no tunability and must be the mirror image of a fixed source,
otherwise the intensity will decrease drastically \cite{zhang2017}. In fact,
there is a hidden parameter dependence, $\mathcal{G}\propto1/v_{F}^{2}$, which
clearly shows the enhancement of the focusing intensity by decreasing the
Fermi velocity. If the Fermi velocity can be manipulated, this should be a
more effective way than controling the doping level to enhance the focusing
intensity since GF has the dependence $\mathcal{G}\propto
V_{0}^{0}/v_{F}^{2}$. Due to rapid advances in materials science, many Dirac materials have been discovered with different
Fermi velocities providing various opportunities for electron
optics, e.g., to enhance the focusing intensity. The general Dirac energy
dispersion has two key variables; one is the Fermi velocity and the other
is the anisotropy. The manipulation of Fermi velocity is  promising for
electron optics, which has been investigated previously \cite{PhysRevB.81.075438}, while here we focus on the anisotropy as a new tuning parameter for Veselago focusing.

\section{Veselago focusing  of the anisotropic MDF and its application}

\subsection{Tunable Veselago focusing by the anisotropic MDF}

Using the projection
relations between the isotropic and anisotropic MDFs, the propagation
properties of the anisotropic MDF across PNJ can be obtained (see
Appendix A). Here, we concentrate on Veselago focusing of the anisotropic MDF and
highlight its tunable features. The necessary conditions for the Veselago focusing of the anisotropic MDF in PNJ can be given by using
the following projection relations:

\begin{equation}
\mathbf{r}_{2}=-\frac{1}{\gamma}\mathbf{r}_{1},
 \varepsilon_{F}=\frac{1-\gamma}{1+\gamma}V_{0}.
  \label{PR_PE}%
\end{equation}

Here, $\mathbf{r}_{1}=\mathbf{r}_{1}^0/\Gamma_{N}=(-a/\Gamma_{N},0)$,\ the equation for $\varepsilon_{F}$
is from Eq. (\ref{SRED}) and $\varepsilon_{N}^{0}=\varepsilon_{P}^{0}$ for the
symmetric PNJ based on isotropic MDF, and $\gamma=\Gamma_{P}/\Gamma_{N}$ is the
ratio of anisotropy of P and N regions. In light of the anisotropy, we can consider three cases; (1) If $\gamma=1$ and $\Gamma_{N}=\Gamma
_{P}=1$,\ it recovers the case for the isotropic MDF, i.e., the Veselago
focusing occurs with mirror symmetry. (2) If $\gamma=1$ and
$\Gamma_{N}=\Gamma_{P}\neq1$, it is for anisotropic MDF. Comparing to case
(1), the inter-site distance for Veselago focusing can be tuned by the degree of
anisotropy, although mirror symmetry is still required. (3) If
$\gamma\neq1$, we also have the anisotropic MDF. In this case, we can tune
the focusing position for a fixing source, and Veselago focusing occurs
in the asymmetric PNJ in contrast to the previous two cases. For Veselago
focusing of the anisotropic MDF, we perform a numerical calculation to show the tunable focusing position by using
$\gamma=2$ and $\gamma=1/2$ in Figs. \ref{focusing}(a) and (c), while Fig.
\ref{focusing}(b) for the isotropic MDF (namely$\ \gamma=1\ $and $\Gamma
_{N}=\Gamma_{P}=1)$\ is used as a reference.

Furthermore, the intensity of\ Veselago focusing can also be tuned by changing
the anisotropy of the MDF. By using the projection relation, the PNJ GF\ based on
the anisotropic MDF can be expressed as $\mathbf{G}(-\mathbf{r}_{1}%
/\gamma,\mathbf{r}_{1},\varepsilon_{F},V_{0})=2/(\Gamma_{N}+\Gamma
_{P})\mathcal{G}(V_{0})$, where $\mathcal{G}$ given by Eq. (\ref{IGF}) is the GF of the isotropic MDF with the
doping level $V_0$, and the prefactor is introduced by
anisotropy of the MDF. Therefore, identical to the isotropic MDF, the focusing
intensity of the anisotropic MDF can also be enhanced by increasing $V_{0}$ and
by decreasing the Fermi velocity $v_{F}$. Because of the ratio
\begin{equation}
\mathbf{G/}\mathcal{G}=2/(\Gamma_{N}+\Gamma_{P}), \label{GFIM}%
\end{equation}
we have the intensity modulation through the anisotropy $\Gamma_{N}$ and
$\Gamma_{P}$\ of the MDF in the N and P regions. The modulation of the focusing
intensity can be clearly seen in Fig. \ref{structure}(d), which compares the
Veselago focusing by considering different values of $\gamma$, and the
quantitative relations among the different intensities are fully described by Eq. (\ref{GFIM}).

\begin{figure}[htp]
\includegraphics[width=0.9\columnwidth,clip]{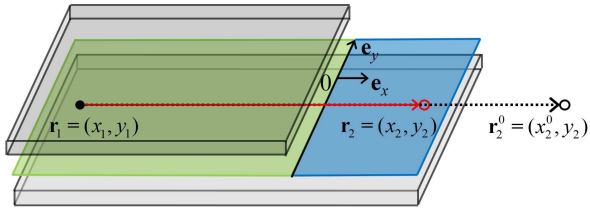}\caption{The proposed
device to probe the masked atom-scale defect in the Cartesian coordinate
system ($\mathbf{{e}}_{x}$, $\mathbf{{e}}_{y}$). Graphene is sandwiched between
the top and bottom substrates, which may contact the gates \cite{LeeNatPhys2015,ChenScience2016}. By using the gates,
one can form the PNJ, which has the left N or encapsulated region with green and the right P or the unencapsulated region with blue. By applying a
superlattice potential or strain on the substrate below the
unencapsulated region, the anisotropic MDF is obtained in the P region with a tunable
degree of anisotropy $\Gamma_{P}$. Assuming a defect denoted by the black dot at $\mathbf{r}%
_{1}=(x_{1},y_{1})$ in the N region, combining the Veselago focusing and
the anisotropic MDF of the P region, the focusing image will occur
at $\mathbf{r}_{2}=(x_{2},y_{2})=(-x_{1}/\Gamma_{P},y_{1})$ (see Eq.
(\ref{PR_PE}) with $\Gamma_{P}>1$) as the red circle or at $\mathbf{r}_{2}%
^{0}=(x_{2}^{0},y_{2})=(x_{1},y_{1})$ (see Eq. (\ref{PR_PE}) with $\Gamma_{P}%
=1$) as the black circle, i.e., the realization of the probe of the masked defect.
Here, an isotropic MDF is assumed in the N region, i.e., $\Gamma_{N}=1$.}%
\label{design}%
\end{figure}

\subsection{Potential applications and discussions}

In the ballistic regime, even a single scatterer may influence the whole
device. A detailed understanding of the influence of such defects on
electronic transport is necessary in order to exploit or avoid their influence
\cite{SettnesPRL2014}. However, it is very difficult to identify masked defects. As a novel application, we propose a device by
utilizing the tunable focusing position of the anisotropic MDF to probe
masked atom-scale defects in two-dimensional materials with graphene as an
example.

To achieve high mobility, it is necessary to encapsulate graphene with insulating and atomically flat boron nitride
crystals. The mismatch
between graphene and the boron nitride crystals usually
brings a small amount of defects into the graphene samples \cite{DeanYoungMericEtAl2010}. Fig. \ref{design} schematically shows the proposed device in which an
incomplete encapsulation is proposed, i.e., graphene is sandwiched between two
substrates and the area of the bottom substrate
is larger than that of the top one. Then,
the encapsulated region and the unencapsulated region can be doped into N type
and P type through the gates contacting the top and bottom substrates, i.e., a
PNJ is formed. Large-area ballistic graphene is not easy to fabricate\cite{LeeNatPhys2015,ChenScience2016}, so in order
to fully utilize the ballistic nature, the encapsulated graphene should be as
large as possible, which leads to a small unencapsulated region for the
probe. To apply the superlattice potential \cite{Park2008,PhysRevLett.101.126804,PhysRevLett.103.046808,PhysRevLett.103.046809,PhysRevB.77.115446,PhysRevB.79.155402,PhysRevLett.105.246803} or the strain on the substrate \cite{PereiraPRL2009,Naumis2017}
below the unencapsulated region, one induces the anisotropic MDF in the P
region whose degree of anisotropy $\Gamma_{P}$ can be fine-tuned, e.g., make
$\Gamma_{P}>1$. If there is a defect denoted by the black dot at
$\mathbf{r}_{1}=(x_{1},y_{1})$ in the N region, due to Veselago focusing,
one can probe a focusing image denoted by the red circle at $\mathbf{r}%
_{2}=(x_{2},y_{2})=(-x_{1}/\Gamma_{P},y_{1})$ (see Eq. (\ref{PR_PE})) with the
strong local density of states in the small unencapsulated region by using a
scanning tunneling microscope, i.e., the realization of the probe of the masked
defect. Here, for the sake of simplicity, the isotropic MDF is assumed in the
N region, i.e., $\Gamma_{N}=1$. In the simple case of the PNJ for the
isotropic MDF, the Veselago focusing can also be used as a probe for
masked defects, but the focusing should be at the mirror image and may be
beyond the unencapsulated region, e.g., see the mirror image at $\mathbf{r}%
_{2}^{0}=(x_{2}^{0},y_{2})=(-x_{1},y_{1})$. Therefore, the tunable focusing
position of anisotropic MDF is beneficial to probe masked defects.

Subsequently, we discuss the experimental feasibility of Veselago focusing of the anisotropic MDF. On the one hand, in order to analytically reveal the underlying physics that is generally applicable to various Dirac materials hosting anisotropic MDFs, we consider the PNJ with an abrupt change of anisotropy and on-site potential in our model study. The Veselago focusing is determined by the propagation phase of the MDF between the source and the probe (see Eq. (A2)). Since the MDF propagates mainly in the uniform regions of PNJ, the Veselago focusing should also exist in the presence of a smooth region for the anisotropy and on-site potential. The same physics can also explain the experimental verification of Veselago focusing in the PNJ with a smooth potential region \cite{LeeNatPhys2015,ChenScience2016}. To the specific material, e.g., strained graphene whose anisotropy is highly tunable by an elastic or piezoelectric substrate \cite{RafaelJPCM2015}, the quantitative atomic simulation can be performed by a proper numerical method \cite{ZhangPRB2017} and is necessary for comparison with future experiments. On the other hand, the realization of Veselago focusing requires high mobility samples. Fortunately, the continuing advances of experimental technology allow the fabrication of high mobility Dirac materials hosting anisotropic MDFs, e.g., graphene with a superlattice potential \cite{171001365F}, ZrTe$_5$ \cite{YuanNPG2016,NL6b02629} and Cd$_3$As$_2$ \cite{NeupaneNC2014} which are inherently anisotropic. Furthermore, in order to observe Veselago focusing of anisotropic MDF in PNJs based on different Dirac materials, the tunable doping is essential and should not severely reduce sample mobility. Therefore, the electrical gating, which is widely used for two-dimensional systems \cite{LeeNatPhys2015,ChenScience2016}, is a better way to dope the sample since the chemical doping may greatly decrease the quality of the sample. In addition, we note that the Veselago focusing of anisotropic MDF could be verified in artificial graphene of cold atoms in light of the recent demonstration of Veselago focusing of isotropic MDFs \cite{LederNC2014} and the tunable dispersion properties of cold atoms in an optical lattice \cite{TarruellN2012}.

We have shown clearly the Veselago focusing of the anisotropic MDF and its tunability, which will offer easy access to future theoretical and experimental studies. Such Veselago focusing has many potential
applications \cite{CheianovScience2007,PhysRevLett.100.236801,MoghaddamPRL2010,SilveirinhaPRL2013,ZhaoPRL2013,ncomms15783,zhang2017,PhysRevB.95.214103}. Since Veselago focusing in the PNJ based on the
anisotropic MDF shows superior tunable features, this must also favor applications. It is convenient to expand our study to incorporate other degrees of freedom such as spin \cite{MoghaddamPRL2010,ZhaoPRL2013,zhang2017} and valley \cite{PhysRevLett.100.236801}, to consider a three-dimensional MDF \cite{PhysRevB.95.214103}, and to examine multiple junctions \cite{PhysRevLett.100.236801} and even superlattices \cite{SilveirinhaPRL2013}. Therefore, this study paves the way for an investigation of electron optics behavior of anisotropic
MDFs with potential device applications.

\section{Conclusions}

In this study, we investigated the propagation of anisotropic MDF in a PNJ structure. We constructed projection relations tuning the anisotropic MDF into isotropic MDF. We analytically showed the precise Veselago focusing and stressed its tunable features which are favorable for the design of novel devices to probe (e.g., masked defects) by utilizing the tunable focusing position. This study presents an innovative concept to realize tunable Veselago focusing, and it paves the way for an investigation of electron optics of anisotropic MDF.

\section*{Acknowledgements}

This work was supported by the National Key
R$\&$D Program of China (Grant No. 2017YFA0303400), the NSFC (Grants No.
11504018, and No. 11774021), the MOST of China (Grants
No. 2014CB848700), and the NSFC program for ``Scientific Research
Center'' (Grant No. U1530401). Support by the bilateral project (FWO-MOST) is gratefully acknowledged. S.H.Z. is
also supported by "the Fundamental Research Funds for the Central Universities
(ZY1824)". We acknowledge the computational
support from the Beijing Computational Science Research Center (CSRC).

\appendix

\section{Analytical derivation of the PNJ GF for
isotropic MDF}

Here, we present a detailed analytical derivation of the PNJ GF for
isotropic MDF, which helps to construct an intuitive physical picture for the propagation
properties of isotropic and anisotropic MDF across the PNJ. The PNJ GF of the
isotropic MDF is \cite{zhang2017},%

\begin{equation}
\mathbf{G}^{0}=\int\frac{dk_{y}}{2\pi}e^{i\phi^{0}(k_{y})}t^{0}(\varphi
_{N}^{0},\varphi_{P}^{0})\frac{|u_{-}(-k_{P,x}^{0},k_{y})\rangle\langle
u_{+}(k_{N,x}^{0},k_{y})|}{iv_{F}k_{N,x}^{0}/k_{N}^{0}}%
\end{equation}
where $k_{N}^{0}\equiv\varepsilon_{N}^{0}/v_{F}$, $k_{P}^{0}\equiv
\varepsilon_{P}^{0}/v_{F}$, $k_{N,x}^{0}\equiv\sqrt{(k_{N}^{0})^{2}-k_{y}^{2}%
}$, $k_{P,x}^{0}\equiv\sqrt{(k_{P}^{0})^{2}-k_{y}^{2}}$, and the propagation
phase is%

\begin{equation}
\phi^{0}(k_{y})=k_{y}Y+a\sqrt{(k_{N}^{0})^{2}-k_{y}^{2}}-X\sqrt{(k_{P}%
^{0})^{2}-k_{y}^{2}}.
\end{equation}

Here, for the sake of simplicity, we assume $\mathbf{r}_{1}^{0}=(-a,0)$ and
$\mathbf{r}_{2}^{0}=(X,Y)$. In polar coordinates, we use $\varphi_{N}%
^{0}\in\lbrack-\pi/2,\pi/2]$ for the incident angle and $\varphi_{P}^{0}%
\in\lbrack-\pi/2,\pi/2]$ for the refractive angle, as defined by $k_{N,x}%
^{0}=k_{N}^{0}\cos\varphi_{N}^{0}$ and $k_{P,x}^{0}=k_{P}^{0}\cos\varphi
_{P}^{0}$. Then $\varphi_{N}^{0}$ and $\varphi_{P}^{0}$ are connected via
$k_{y}=k_{N}^{0}\sin\varphi_{N}^{0}=k_{P}^{0}\sin\varphi_{P}^{0}$ and the
transmission coefficient is
\begin{equation}
t^{0}(\varphi_{N}^{0},\varphi_{P}^{0})=\frac{2\cos\varphi_{N}^{0}%
}{e^{-i\varphi_{N}^{0}}+{e}^{-i\varphi_{P}^{0}}}%
\end{equation}
corresponding to the propagation of the isotropic MDF from the left N region\ to
the right P region. $t^{0}(\varphi_{N}^{0},\varphi_{P}^{0})$ implies high
transparency of PNJ based on the isotropic MDF
\cite{KatsnelsonNatPhys2006,CheianovPRB2006}, which is an important factor
beneficial for the realization of Veselago focusing \cite{CheianovScience2007}.

The isotropic MDF across the PNJ exhibits novel electron optics behaviors
similar to those in metamaterials with a negative refraction index, e.g.,
Veselago focusing and caustics \cite{CheianovScience2007}, which can be
explained by examining the classical trajectory determined by the propagation
phase. The classical trajectory $k_{y,\mathrm{c}}$ going from $\mathbf{r}%
_{1}^{0}$ to $\mathbf{r}_{2}^{0}$ is determined by $\partial\phi^{0}%
(k_{y})/\partial k_{y}=0$ as%

\begin{equation}
Y=a\tan\theta_{N}^{0}-X\tan\theta_{P}^{0}=R_{N}^{0}\sin\theta_{N}^{0}%
-R_{P}^{0}\sin\theta_{P}^{0}, \label{CT}%
\end{equation}
where we have introduced the classical incident angle $\theta_{N}^{0}%
\in\lbrack-\pi/2,\pi/2]$ and the refractive angle $\theta_{P}^{0}\in\lbrack
-\pi/2,\pi/2]$ through%
\begin{align}
\tan\theta_{N}^{0}  &  \equiv\frac{k_{y,\mathrm{c}}}{\sqrt{(k_{N}^{0}%
)^{2}-k_{y,\mathrm{c}}^{2}}},\\
\tan\theta_{P}^{0}  &  \equiv\frac{k_{y,\mathrm{c}}}{\sqrt{(k_{P}^{0}%
)^{2}-k_{y,\mathrm{c}}^{2}}},
\end{align}
and the classical path in the N region and the P region, i.e., $R_{N}^{0}\equiv
a/\cos\theta_{N}^{0}$ and $R_{P}^{0}\equiv X/\cos\theta_{P}^{0}$.
From
$\sin\theta_{N}^{0}=k_{y,\mathrm{c}}/k_{N}^{0}$ and $\sin\theta_{P}%
^{0}=k_{y,\mathrm{c}}/k_{P}^{0}$, we further have
\begin{equation}
\frac{\sin\theta_{N}^{0}}{\sin\theta_{P}^{0}}=\frac{k_{P}^{0}}{k_{N}^{0}%
}=\frac{\varepsilon_{P}^{0}}{\varepsilon_{N}^{0}}\equiv n^{0}, \label{MM}%
\end{equation}
which, together with Eq. (\ref{CT}), completely determines $\theta_{N}%
^{0},\theta_{P}^{0}$ and hence the classical trajectory. Here, the definition
of $\theta_{P}^{0}$ makes $n^{0}$ be the magnitude of the effective refractive
index of the PNJ, which should be negative \cite{CheianovScience2007}.

\subsection{Veselago focusing and the analytical Green's function}

For the
symmetric PNJ, we have $k_{N}^{0}=k_{P}^{0}\equiv k_{F}^{0}$, $\theta_{N}%
^{0}=\theta_{P}^{0}\equiv\theta^{0}$, and then the classical trajectory is%

\begin{equation}
Y=\tan\theta^{0}(a-X)=(R_{N}^{0}-R_{P}^{0})\sin\theta^{0}.
\end{equation}

Along the classical path, the phase is
\begin{equation}
\Phi^{0}=k_{F}^{0}\frac{a-X}{\cos\theta^{0}}=\frac{k_{F}^{0}Y}{\sin\theta^{0}%
}=k_{F}^{0}(R_{N}^{0}-R_{P}^{0})=\mathrm{sgn}(a-X)k_{F}^{0}|\mathbf{r}_{2}%
^{0}-\mathbf{r}_{1\mathrm{m}}^{0}|,
\end{equation}
where $\mathbf{r}_{1\mathrm{m}}^{0}=(a,0)$ is the mirror image of
$\mathbf{r}_{1}^{0}=(-a,0)$. In particular, if $\mathbf{r}_{2}%
^{0}=\mathbf{r}_{1\mathrm{m}}^{0}$, the $0$-order term $\Phi^{0}$\ and the
arbitrary-order derivative of $\phi^{0}(k_{y})$ vanish for all classical
trajectories with different $k_{y,\mathrm{c}}$, which leads to the Veselago
focusing.

To derive the analytical GF for the symmetric PNJ, we define the incident
angle $\varphi^{0}\in\lbrack-\pi/2,\pi/2]$ through $k_{x}^{0}=k_{F}^{0}%
\cos\varphi^{0}$ and $k_{y}=k_{F}^{0}\sin\varphi^{0}$. Then using the
transmission coefficient $t^{0}=\cos\varphi^{0}e^{i\varphi^{0}}$, we have
(keeping traveling waves only):%

\begin{align}
\mathbf{G}^{0}  &  \approx\frac{k_{F}^{0}}{2\pi iv_{F}}\int_{-\pi/2}^{\pi
/2}\cos\varphi^{0}e^{i\varphi^{0}}d\varphi^{0}\ e^{i\phi^{0}}|u_{-}^{0}%
(-k_{x}^{0},k_{y})\rangle\langle u_{+}^{0}(k_{x}^{0},k_{y})|\\
&  =\frac{k_{F}^{0}}{4\pi iv_{F}}(g_{0}+g_{y}\sigma_{x}+g_{z}\sigma_{z}),
\end{align}
where%

\begin{align}
g_{0}^{0}  &  \equiv\int_{-\pi/2}^{\pi/2}\cos^{2}\varphi^{0}e^{i\phi^{0}%
}d\varphi^{0},\\
g_{y}^{0}  &  \equiv\int_{-\pi/2}^{\pi/2}\cos\varphi^{0}e^{i\phi^{0}}%
d\varphi^{0},\\
g_{z}^{0}  &  \equiv i\int_{-\pi/2}^{\pi/2}\sin\varphi^{0}\cos\varphi
^{0}e^{i\phi^{0}}d\varphi^{0}.
\end{align}
Here, the propagation phase
\begin{equation}
\phi^{0}(\varphi^{0})=k_{F}^{0}[Y\sin\varphi^{0}+(a-X)\cos\varphi^{0}%
]=\Phi\cos(\varphi^{0}-\theta^{0})
\end{equation}
and $\theta^{0}=\tan^{-1}[Y/(a-X)]$ is the classical incident angle and
$\Phi^{0}\equiv k_{F}^{0}(a-X)/\cos\theta^{0}$ is the phase along the
classical trajectory. For $X=a$ and $Y=0$, i.e., $\mathbf{r}_{2}%
^{0}=\mathbf{r}_{1\mathrm{m}}^{0}$, we have $\phi^{0}=0$ and hence $g_{0}%
^{0}=\pi/2$, $g_{y}^{0}=2$, $g_{z}^{0}=0$, i.e.,
\begin{equation}
\mathbf{G}^{0}(\mathbf{r}_{2}^{0},\mathbf{r}_{1}^{0},\varepsilon_{F}^{0}%
,V_{0}^{0})=\frac{k_{F}^{0}}{4\pi iv_{F}}(\frac{\pi}{2}+2\sigma_{x}).
\label{GFI}%
\end{equation}

The analytical expression for the GF shows the dependence on the material parameters
such as $v_{F}$, $\varepsilon_{F}^{0}$ and $V_{0}^{0}$, but it does not depend on
the position vectors $\mathbf{r}_{1}^{0}$ and $\mathbf{r}_{2}^{0}$ as long
as\textbf{ }$\mathbf{r}_{2}^{0}=\mathbf{r}_{1\mathrm{m}}^{0}$.

\subsection{Caustics}

For the asymmetric PNJ, there is the caustics
corresponding to the singularity of the classical trajectory. For
$\mathbf{R}_{2}=(X,Y)\ $at certain special locations, the quadratic term of
$\phi^{0}(k_{y})$ also vanishes, i.e., $w^{0}\equiv\partial\lbrack\phi
^{0}(k_{y})]^{2}/\partial k_{y}^{2}=0$ with
\begin{equation}
w^{0}=\frac{a}{q_{N}^{0}\cos^{3}\theta_{N}^{0}}-\frac{X}{q_{P}^{0}\cos
^{3}\theta_{P}^{0}}.
\end{equation}

Therefore, $w^{0}=0$ leads to the equation%

\begin{equation}
\frac{\cos\theta_{N}^{0}}{\cos\theta_{P}^{0}}=(\frac{aq_{P}^{0}}{Xq_{N}^{0}%
})^{1/3}=(\frac{n^{0}a}{X})^{1/3}. \label{HO}%
\end{equation}

The caustics curve is determined by Eqs.\ (\ref{CT}), (\ref{MM}), and
(\ref{HO}):
\begin{align}
Y  &  =\pm\frac{\left[  X_{cusp}^{2/3}-X^{2/3}\right]  ^{3/2}}{\sqrt
{1-(n^{0})^{2}}}\ (\mathrm{for}\ n^{0}<1\ \mathrm{and\ }X<X_{cusp}),\\
Y  &  =\pm\frac{\left[  X^{2/3}-X_{cusp}^{2/3}\right]  ^{3/2}}{\sqrt
{(n^{0})^{2}-1}}\ (\mathrm{for}\ n^{0}>1\ \mathrm{and\ }X>X_{cusp}).
\end{align}

Note that there is no solution for $n^{0}<1$, $X>X_{cusp}$; and $n^{0}%
>1,X<X_{cusp}$. Here, $X_{cusp}=n^{0}a$ is the position of the cusp which is a
singularity in the density of classical trajectories.


\begin{thebibliography}{56}
\expandafter\ifx\csname natexlab\endcsname\relax\def\natexlab#1{#1}\fi
\expandafter\ifx\csname bibnamefont\endcsname\relax
  \def\bibnamefont#1{#1}\fi
\expandafter\ifx\csname bibfnamefont\endcsname\relax
  \def\bibfnamefont#1{#1}\fi
\expandafter\ifx\csname citenamefont\endcsname\relax
  \def\citenamefont#1{#1}\fi
\expandafter\ifx\csname url\endcsname\relax
  \def\url#1{\texttt{#1}}\fi
\expandafter\ifx\csname urlprefix\endcsname\relax\def\urlprefix{URL }\fi
\providecommand{\bibinfo}[2]{#2}
\providecommand{\eprint}[2][]{\url{#2}}

\bibitem[{\citenamefont{Castro~Neto et~al.}(2009)\citenamefont{Castro~Neto,
  Guinea, Peres, Novoselov, and Geim}}]{CastroRMP2009}
\bibinfo{author}{\bibfnamefont{A.~H.} \bibnamefont{Castro~Neto}},
  \bibinfo{author}{\bibfnamefont{F.}~\bibnamefont{Guinea}},
  \bibinfo{author}{\bibfnamefont{N.~M.~R.} \bibnamefont{Peres}},
  \bibinfo{author}{\bibfnamefont{K.~S.} \bibnamefont{Novoselov}},
  \bibnamefont{and} \bibinfo{author}{\bibfnamefont{A.~K.} \bibnamefont{Geim}},
  \bibinfo{journal}{Rev. Mod. Phys.} \textbf{\bibinfo{volume}{81}},
  \bibinfo{pages}{109} (\bibinfo{year}{2009}).

\bibitem[{\citenamefont{Qi and Zhang}(2011)}]{QiRMP2011}
\bibinfo{author}{\bibfnamefont{X.-L.} \bibnamefont{Qi}} \bibnamefont{and}
  \bibinfo{author}{\bibfnamefont{S.-C.} \bibnamefont{Zhang}},
  \bibinfo{journal}{Rev. Mod. Phys.} \textbf{\bibinfo{volume}{83}},
  \bibinfo{pages}{1057} (\bibinfo{year}{2011}).

\bibitem[{\citenamefont{Wehling et~al.}(2014)\citenamefont{Wehling,
  Black-Schaffer, and Balatsky}}]{Wehling2014}
\bibinfo{author}{\bibfnamefont{T.}~\bibnamefont{Wehling}},
  \bibinfo{author}{\bibfnamefont{A.}~\bibnamefont{Black-Schaffer}},
  \bibnamefont{and} \bibinfo{author}{\bibfnamefont{A.}~\bibnamefont{Balatsky}},
  \bibinfo{journal}{Advances in Physics} \textbf{\bibinfo{volume}{63}},
  \bibinfo{pages}{1} (\bibinfo{year}{2014}).

\bibitem[{\citenamefont{Wang et~al.}(2015)\citenamefont{Wang, Deng, Liu, and
  Liu}}]{WangNSRl2015}
\bibinfo{author}{\bibfnamefont{J.}~\bibnamefont{Wang}},
  \bibinfo{author}{\bibfnamefont{S.}~\bibnamefont{Deng}},
  \bibinfo{author}{\bibfnamefont{Z.}~\bibnamefont{Liu}}, \bibnamefont{and}
  \bibinfo{author}{\bibfnamefont{Z.}~\bibnamefont{Liu}},
  \bibinfo{journal}{National Science Review} \textbf{\bibinfo{volume}{2}},
  \bibinfo{pages}{22} (\bibinfo{year}{2015}).

\bibitem[{\citenamefont{Kobayashi et~al.}(2007)\citenamefont{Kobayashi,
  Katayama, Suzumura, and Fukuyama}}]{Kobayashi2007}
\bibinfo{author}{\bibfnamefont{A.}~\bibnamefont{Kobayashi}},
  \bibinfo{author}{\bibfnamefont{S.}~\bibnamefont{Katayama}},
  \bibinfo{author}{\bibfnamefont{Y.}~\bibnamefont{Suzumura}}, \bibnamefont{and}
  \bibinfo{author}{\bibfnamefont{H.}~\bibnamefont{Fukuyama}},
  \bibinfo{journal}{Journal of the Physical Society of Japan}
  \textbf{\bibinfo{volume}{76}}, \bibinfo{pages}{034711}
  (\bibinfo{year}{2007}).

\bibitem[{\citenamefont{Banerjee et~al.}(2009)\citenamefont{Banerjee, Singh,
  Pardo, and Pickett}}]{PhysRevLett.103.016402}
\bibinfo{author}{\bibfnamefont{S.}~\bibnamefont{Banerjee}},
  \bibinfo{author}{\bibfnamefont{R.~R.~P.} \bibnamefont{Singh}},
  \bibinfo{author}{\bibfnamefont{V.}~\bibnamefont{Pardo}}, \bibnamefont{and}
  \bibinfo{author}{\bibfnamefont{W.~E.} \bibnamefont{Pickett}},
  \bibinfo{journal}{Phys. Rev. Lett.} \textbf{\bibinfo{volume}{103}},
  \bibinfo{pages}{016402} (\bibinfo{year}{2009}).

\bibitem[{\citenamefont{Richard et~al.}(2010)\citenamefont{Richard, Nakayama,
  Sato, Neupane, Xu, Bowen, Chen, Luo, Wang, Dai
  et~al.}}]{PhysRevLett.104.137001}
\bibinfo{author}{\bibfnamefont{P.}~\bibnamefont{Richard}},
  \bibinfo{author}{\bibfnamefont{K.}~\bibnamefont{Nakayama}},
  \bibinfo{author}{\bibfnamefont{T.}~\bibnamefont{Sato}},
  \bibinfo{author}{\bibfnamefont{M.}~\bibnamefont{Neupane}},
  \bibinfo{author}{\bibfnamefont{Y.-M.} \bibnamefont{Xu}},
  \bibinfo{author}{\bibfnamefont{J.~H.} \bibnamefont{Bowen}},
  \bibinfo{author}{\bibfnamefont{G.~F.} \bibnamefont{Chen}},
  \bibinfo{author}{\bibfnamefont{J.~L.} \bibnamefont{Luo}},
  \bibinfo{author}{\bibfnamefont{N.~L.} \bibnamefont{Wang}},
  \bibinfo{author}{\bibfnamefont{X.}~\bibnamefont{Dai}}, \bibnamefont{et~al.},
  \bibinfo{journal}{Phys. Rev. Lett.} \textbf{\bibinfo{volume}{104}},
  \bibinfo{pages}{137001} (\bibinfo{year}{2010}).

\bibitem[{\citenamefont{Killi et~al.}(2011)\citenamefont{Killi, Wu, and
  Paramekanti}}]{PhysRevLett.107.086801}
\bibinfo{author}{\bibfnamefont{M.}~\bibnamefont{Killi}},
  \bibinfo{author}{\bibfnamefont{S.}~\bibnamefont{Wu}}, \bibnamefont{and}
  \bibinfo{author}{\bibfnamefont{A.}~\bibnamefont{Paramekanti}},
  \bibinfo{journal}{Phys. Rev. Lett.} \textbf{\bibinfo{volume}{107}},
  \bibinfo{pages}{086801} (\bibinfo{year}{2011}).

\bibitem[{\citenamefont{Jo et~al.}(2014)\citenamefont{Jo, Park, Lee, Eom, Choi,
  Shim, Kang, and Kim}}]{PhysRevLett.113.156602}
\bibinfo{author}{\bibfnamefont{Y.~J.} \bibnamefont{Jo}},
  \bibinfo{author}{\bibfnamefont{J.}~\bibnamefont{Park}},
  \bibinfo{author}{\bibfnamefont{G.}~\bibnamefont{Lee}},
  \bibinfo{author}{\bibfnamefont{M.~J.} \bibnamefont{Eom}},
  \bibinfo{author}{\bibfnamefont{E.~S.} \bibnamefont{Choi}},
  \bibinfo{author}{\bibfnamefont{J.~H.} \bibnamefont{Shim}},
  \bibinfo{author}{\bibfnamefont{W.}~\bibnamefont{Kang}}, \bibnamefont{and}
  \bibinfo{author}{\bibfnamefont{J.~S.} \bibnamefont{Kim}},
  \bibinfo{journal}{Phys. Rev. Lett.} \textbf{\bibinfo{volume}{113}},
  \bibinfo{pages}{156602} (\bibinfo{year}{2014}).

\bibitem[{\citenamefont{Moon et~al.}(2011)\citenamefont{Moon, Han, Lee, and
  Choi}}]{PhysRevB.84.195425}
\bibinfo{author}{\bibfnamefont{C.-Y.} \bibnamefont{Moon}},
  \bibinfo{author}{\bibfnamefont{J.}~\bibnamefont{Han}},
  \bibinfo{author}{\bibfnamefont{H.}~\bibnamefont{Lee}}, \bibnamefont{and}
  \bibinfo{author}{\bibfnamefont{H.~J.} \bibnamefont{Choi}},
  \bibinfo{journal}{Phys. Rev. B} \textbf{\bibinfo{volume}{84}},
  \bibinfo{pages}{195425} (\bibinfo{year}{2011}).

\bibitem[{\citenamefont{Lopez-Bezanilla and
  Littlewood}(2016)}]{PhysRevB.93.241405}
\bibinfo{author}{\bibfnamefont{A.}~\bibnamefont{Lopez-Bezanilla}}
  \bibnamefont{and} \bibinfo{author}{\bibfnamefont{P.~B.}
  \bibnamefont{Littlewood}}, \bibinfo{journal}{Phys. Rev. B}
  \textbf{\bibinfo{volume}{93}}, \bibinfo{pages}{241405}
  (\bibinfo{year}{2016}).

\bibitem[{\citenamefont{Li et~al.}(2017)\citenamefont{Li, Cao, Wu, and
  Louie}}]{li2017}
\bibinfo{author}{\bibfnamefont{Z.}~\bibnamefont{Li}},
  \bibinfo{author}{\bibfnamefont{T.}~\bibnamefont{Cao}},
  \bibinfo{author}{\bibfnamefont{M.}~\bibnamefont{Wu}}, \bibnamefont{and}
  \bibinfo{author}{\bibfnamefont{S.~G.} \bibnamefont{Louie}},
  \bibinfo{journal}{Nano Letters} \textbf{\bibinfo{volume}{17}},
  \bibinfo{pages}{2280} (\bibinfo{year}{2017}).

\bibitem[{\citenamefont{Park et~al.}(2011)\citenamefont{Park, Lee,
  Wolff-Fabris, Koh, Eom, Kim, Farhan, Jo, Kim, Shim
  et~al.}}]{PhysRevLett.107.126402}
\bibinfo{author}{\bibfnamefont{J.}~\bibnamefont{Park}},
  \bibinfo{author}{\bibfnamefont{G.}~\bibnamefont{Lee}},
  \bibinfo{author}{\bibfnamefont{F.}~\bibnamefont{Wolff-Fabris}},
  \bibinfo{author}{\bibfnamefont{Y.~Y.} \bibnamefont{Koh}},
  \bibinfo{author}{\bibfnamefont{M.~J.} \bibnamefont{Eom}},
  \bibinfo{author}{\bibfnamefont{Y.~K.} \bibnamefont{Kim}},
  \bibinfo{author}{\bibfnamefont{M.~A.} \bibnamefont{Farhan}},
  \bibinfo{author}{\bibfnamefont{Y.~J.} \bibnamefont{Jo}},
  \bibinfo{author}{\bibfnamefont{C.}~\bibnamefont{Kim}},
  \bibinfo{author}{\bibfnamefont{J.~H.} \bibnamefont{Shim}},
  \bibnamefont{et~al.}, \bibinfo{journal}{Phys. Rev. Lett.}
  \textbf{\bibinfo{volume}{107}}, \bibinfo{pages}{126402}
  (\bibinfo{year}{2011}).

\bibitem[{\citenamefont{Mullen et~al.}(2015)\citenamefont{Mullen, Uchoa, and
  Glatzhofer}}]{PhysRevLett.115.026403}
\bibinfo{author}{\bibfnamefont{K.}~\bibnamefont{Mullen}},
  \bibinfo{author}{\bibfnamefont{B.}~\bibnamefont{Uchoa}}, \bibnamefont{and}
  \bibinfo{author}{\bibfnamefont{D.~T.} \bibnamefont{Glatzhofer}},
  \bibinfo{journal}{Phys. Rev. Lett.} \textbf{\bibinfo{volume}{115}},
  \bibinfo{pages}{026403} (\bibinfo{year}{2015}).

\bibitem[{\citenamefont{Yan et~al.}(2017)\citenamefont{Yan, Huang, Zhang, Wang,
  Yao, Deng, Wan, Zhang, Arita, Yang et~al.}}]{Yan2017}
\bibinfo{author}{\bibfnamefont{M.}~\bibnamefont{Yan}},
  \bibinfo{author}{\bibfnamefont{H.}~\bibnamefont{Huang}},
  \bibinfo{author}{\bibfnamefont{K.}~\bibnamefont{Zhang}},
  \bibinfo{author}{\bibfnamefont{E.}~\bibnamefont{Wang}},
  \bibinfo{author}{\bibfnamefont{W.}~\bibnamefont{Yao}},
  \bibinfo{author}{\bibfnamefont{K.}~\bibnamefont{Deng}},
  \bibinfo{author}{\bibfnamefont{G.}~\bibnamefont{Wan}},
  \bibinfo{author}{\bibfnamefont{H.}~\bibnamefont{Zhang}},
  \bibinfo{author}{\bibfnamefont{M.}~\bibnamefont{Arita}},
  \bibinfo{author}{\bibfnamefont{H.}~\bibnamefont{Yang}}, \bibnamefont{et~al.},
  \bibinfo{journal}{Nature Communications} \textbf{\bibinfo{volume}{8}},
  \bibinfo{pages}{257} (\bibinfo{year}{2017}).

\bibitem[{\citenamefont{Pereira and Castro~Neto}(2009)}]{PereiraPRL2009}
\bibinfo{author}{\bibfnamefont{V.~M.} \bibnamefont{Pereira}} \bibnamefont{and}
  \bibinfo{author}{\bibfnamefont{A.~H.} \bibnamefont{Castro~Neto}},
  \bibinfo{journal}{Phys. Rev. Lett.} \textbf{\bibinfo{volume}{103}},
  \bibinfo{pages}{046801} (\bibinfo{year}{2009}).

\bibitem[{\citenamefont{Naumis et~al.}(2017)\citenamefont{Naumis,
  Barraza-Lopez, Oliva-Leyva, and Terrones}}]{Naumis2017}
\bibinfo{author}{\bibfnamefont{G.~G.} \bibnamefont{Naumis}},
  \bibinfo{author}{\bibfnamefont{S.}~\bibnamefont{Barraza-Lopez}},
  \bibinfo{author}{\bibfnamefont{M.}~\bibnamefont{Oliva-Leyva}},
  \bibnamefont{and} \bibinfo{author}{\bibfnamefont{H.}~\bibnamefont{Terrones}},
  \bibinfo{journal}{Reports on Progress in Physics}
  \textbf{\bibinfo{volume}{80}}, \bibinfo{pages}{096501}
  (\bibinfo{year}{2017}).

\bibitem[{\citenamefont{Park et~al.}(2008{\natexlab{a}})\citenamefont{Park,
  Yang, Son, Cohen, and Louie}}]{Park2008}
\bibinfo{author}{\bibfnamefont{C.-H.} \bibnamefont{Park}},
  \bibinfo{author}{\bibfnamefont{L.}~\bibnamefont{Yang}},
  \bibinfo{author}{\bibfnamefont{Y.-W.} \bibnamefont{Son}},
  \bibinfo{author}{\bibfnamefont{M.~L.} \bibnamefont{Cohen}}, \bibnamefont{and}
  \bibinfo{author}{\bibfnamefont{S.~G.} \bibnamefont{Louie}},
  \bibinfo{journal}{Nature Physics} \textbf{\bibinfo{volume}{4}},
  \bibinfo{pages}{213} (\bibinfo{year}{2008}{\natexlab{a}}).

\bibitem[{\citenamefont{Barbier et~al.}(2010)\citenamefont{Barbier,
  Vasilopoulos, and Peeters}}]{PhysRevB.81.075438}
\bibinfo{author}{\bibfnamefont{M.}~\bibnamefont{Barbier}},
  \bibinfo{author}{\bibfnamefont{P.}~\bibnamefont{Vasilopoulos}},
  \bibnamefont{and} \bibinfo{author}{\bibfnamefont{F.~M.}
  \bibnamefont{Peeters}}, \bibinfo{journal}{Phys. Rev. B}
  \textbf{\bibinfo{volume}{81}}, \bibinfo{pages}{075438}
  (\bibinfo{year}{2010}).

\bibitem[{\citenamefont{Park et~al.}(2008{\natexlab{b}})\citenamefont{Park,
  Yang, Son, Cohen, and Louie}}]{PhysRevLett.101.126804}
\bibinfo{author}{\bibfnamefont{C.-H.} \bibnamefont{Park}},
  \bibinfo{author}{\bibfnamefont{L.}~\bibnamefont{Yang}},
  \bibinfo{author}{\bibfnamefont{Y.-W.} \bibnamefont{Son}},
  \bibinfo{author}{\bibfnamefont{M.~L.} \bibnamefont{Cohen}}, \bibnamefont{and}
  \bibinfo{author}{\bibfnamefont{S.~G.} \bibnamefont{Louie}},
  \bibinfo{journal}{Phys. Rev. Lett.} \textbf{\bibinfo{volume}{101}},
  \bibinfo{pages}{126804} (\bibinfo{year}{2008}{\natexlab{b}}).

\bibitem[{\citenamefont{Rusponi et~al.}(2010)\citenamefont{Rusponi, Papagno,
  Moras, Vlaic, Etzkorn, Sheverdyaeva, Pacil\'e, Brune, and
  Carbone}}]{PhysRevLett.105.246803}
\bibinfo{author}{\bibfnamefont{S.}~\bibnamefont{Rusponi}},
  \bibinfo{author}{\bibfnamefont{M.}~\bibnamefont{Papagno}},
  \bibinfo{author}{\bibfnamefont{P.}~\bibnamefont{Moras}},
  \bibinfo{author}{\bibfnamefont{S.}~\bibnamefont{Vlaic}},
  \bibinfo{author}{\bibfnamefont{M.}~\bibnamefont{Etzkorn}},
  \bibinfo{author}{\bibfnamefont{P.~M.} \bibnamefont{Sheverdyaeva}},
  \bibinfo{author}{\bibfnamefont{D.}~\bibnamefont{Pacil\'e}},
  \bibinfo{author}{\bibfnamefont{H.}~\bibnamefont{Brune}}, \bibnamefont{and}
  \bibinfo{author}{\bibfnamefont{C.}~\bibnamefont{Carbone}},
  \bibinfo{journal}{Phys. Rev. Lett.} \textbf{\bibinfo{volume}{105}},
  \bibinfo{pages}{246803} (\bibinfo{year}{2010}).

\bibitem[{\citenamefont{Lu et~al.}(2016)\citenamefont{Lu, Cuamba, Lin, Hao,
  Wang, Li, Zhao, and Ting}}]{PhysRevB.94.195423}
\bibinfo{author}{\bibfnamefont{H.-Y.} \bibnamefont{Lu}},
  \bibinfo{author}{\bibfnamefont{A.~S.} \bibnamefont{Cuamba}},
  \bibinfo{author}{\bibfnamefont{S.-Y.} \bibnamefont{Lin}},
  \bibinfo{author}{\bibfnamefont{L.}~\bibnamefont{Hao}},
  \bibinfo{author}{\bibfnamefont{R.}~\bibnamefont{Wang}},
  \bibinfo{author}{\bibfnamefont{H.}~\bibnamefont{Li}},
  \bibinfo{author}{\bibfnamefont{Y.}~\bibnamefont{Zhao}}, \bibnamefont{and}
  \bibinfo{author}{\bibfnamefont{C.~S.} \bibnamefont{Ting}},
  \bibinfo{journal}{Phys. Rev. B} \textbf{\bibinfo{volume}{94}},
  \bibinfo{pages}{195423} (\bibinfo{year}{2016}).

\bibitem[{\citenamefont{Cheianov et~al.}(2007)\citenamefont{Cheianov, Fal'ko,
  and Altshuler}}]{CheianovScience2007}
\bibinfo{author}{\bibfnamefont{V.~V.} \bibnamefont{Cheianov}},
  \bibinfo{author}{\bibfnamefont{V.}~\bibnamefont{Fal'ko}}, \bibnamefont{and}
  \bibinfo{author}{\bibfnamefont{B.~L.} \bibnamefont{Altshuler}},
  \bibinfo{journal}{Science} \textbf{\bibinfo{volume}{315}},
  \bibinfo{pages}{1252} (\bibinfo{year}{2007}).

\bibitem[{\citenamefont{Williams et~al.}(2011)\citenamefont{Williams, Low,
  Lundstrom, and Marcus}}]{WilliamsNatNano2011}
\bibinfo{author}{\bibfnamefont{J.~R.} \bibnamefont{Williams}},
  \bibinfo{author}{\bibfnamefont{T.}~\bibnamefont{Low}},
  \bibinfo{author}{\bibfnamefont{M.~S.} \bibnamefont{Lundstrom}},
  \bibnamefont{and} \bibinfo{author}{\bibfnamefont{C.~M.}
  \bibnamefont{Marcus}}, \bibinfo{journal}{Nat Nano}
  \textbf{\bibinfo{volume}{6}}, \bibinfo{pages}{222} (\bibinfo{year}{2011}).

\bibitem[{\citenamefont{Rickhaus et~al.}(2013)\citenamefont{Rickhaus, Maurand,
  Liu, Weiss, Richter, and Schonenberger}}]{RickhausNatComm2013}
\bibinfo{author}{\bibfnamefont{P.}~\bibnamefont{Rickhaus}},
  \bibinfo{author}{\bibfnamefont{R.}~\bibnamefont{Maurand}},
  \bibinfo{author}{\bibfnamefont{M.-H.} \bibnamefont{Liu}},
  \bibinfo{author}{\bibfnamefont{M.}~\bibnamefont{Weiss}},
  \bibinfo{author}{\bibfnamefont{K.}~\bibnamefont{Richter}}, \bibnamefont{and}
  \bibinfo{author}{\bibfnamefont{C.}~\bibnamefont{Schonenberger}},
  \bibinfo{journal}{Nat. Commun.} \textbf{\bibinfo{volume}{4}},
  \bibinfo{pages}{2342} (\bibinfo{year}{2013}).

\bibitem[{\citenamefont{Taychatanapat et~al.}(2015)\citenamefont{Taychatanapat,
  Tan, Yeo, Watanabe, Taniguchi, and \"{O}zyilmaz}}]{ncomms7093}
\bibinfo{author}{\bibfnamefont{T.}~\bibnamefont{Taychatanapat}},
  \bibinfo{author}{\bibfnamefont{J.~Y.} \bibnamefont{Tan}},
  \bibinfo{author}{\bibfnamefont{Y.}~\bibnamefont{Yeo}},
  \bibinfo{author}{\bibfnamefont{K.}~\bibnamefont{Watanabe}},
  \bibinfo{author}{\bibfnamefont{T.}~\bibnamefont{Taniguchi}},
  \bibnamefont{and}
  \bibinfo{author}{\bibfnamefont{B.}~\bibnamefont{\"{O}zyilmaz}},
  \bibinfo{journal}{Nature Communications} \textbf{\bibinfo{volume}{6}},
  \bibinfo{pages}{6093} (\bibinfo{year}{2015}).

\bibitem[{\citenamefont{Pendry}(2007)}]{Pendry2007}
\bibinfo{author}{\bibfnamefont{J.~B.} \bibnamefont{Pendry}},
  \bibinfo{journal}{Science} \textbf{\bibinfo{volume}{315}},
  \bibinfo{pages}{1226} (\bibinfo{year}{2007}), ISSN \bibinfo{issn}{0036-8075}.

\bibitem[{\citenamefont{Garcia-Pomar et~al.}(2008)\citenamefont{Garcia-Pomar,
  Cortijo, and Nieto-Vesperinas}}]{PhysRevLett.100.236801}
\bibinfo{author}{\bibfnamefont{J.~L.} \bibnamefont{Garcia-Pomar}},
  \bibinfo{author}{\bibfnamefont{A.}~\bibnamefont{Cortijo}}, \bibnamefont{and}
  \bibinfo{author}{\bibfnamefont{M.}~\bibnamefont{Nieto-Vesperinas}},
  \bibinfo{journal}{Phys. Rev. Lett.} \textbf{\bibinfo{volume}{100}},
  \bibinfo{pages}{236801} (\bibinfo{year}{2008}).

\bibitem[{\citenamefont{Moghaddam and Zareyan}(2010)}]{MoghaddamPRL2010}
\bibinfo{author}{\bibfnamefont{A.~G.} \bibnamefont{Moghaddam}}
  \bibnamefont{and} \bibinfo{author}{\bibfnamefont{M.}~\bibnamefont{Zareyan}},
  \bibinfo{journal}{Phys. Rev. Lett.} \textbf{\bibinfo{volume}{105}},
  \bibinfo{pages}{146803} (\bibinfo{year}{2010}).

\bibitem[{\citenamefont{Silveirinha and Engheta}(2013)}]{SilveirinhaPRL2013}
\bibinfo{author}{\bibfnamefont{M.~G.} \bibnamefont{Silveirinha}}
  \bibnamefont{and} \bibinfo{author}{\bibfnamefont{N.}~\bibnamefont{Engheta}},
  \bibinfo{journal}{Phys. Rev. Lett.} \textbf{\bibinfo{volume}{110}},
  \bibinfo{pages}{213902} (\bibinfo{year}{2013}).

\bibitem[{\citenamefont{Zhao et~al.}(2013)\citenamefont{Zhao, Tang, Gu, and
  Duan}}]{ZhaoPRL2013}
\bibinfo{author}{\bibfnamefont{L.}~\bibnamefont{Zhao}},
  \bibinfo{author}{\bibfnamefont{P.}~\bibnamefont{Tang}},
  \bibinfo{author}{\bibfnamefont{B.-L.} \bibnamefont{Gu}}, \bibnamefont{and}
  \bibinfo{author}{\bibfnamefont{W.}~\bibnamefont{Duan}},
  \bibinfo{journal}{Phys. Rev. Lett.} \textbf{\bibinfo{volume}{111}},
  \bibinfo{pages}{116601} (\bibinfo{year}{2013}).

\bibitem[{\citenamefont{Milovanovic et~al.}(2015)\citenamefont{Milovanovic,
  Moldovan, and Peeters}}]{MilovanovicJAP2015}
\bibinfo{author}{\bibfnamefont{S.~P.} \bibnamefont{Milovanovic}},
  \bibinfo{author}{\bibfnamefont{D.}~\bibnamefont{Moldovan}}, \bibnamefont{and}
  \bibinfo{author}{\bibfnamefont{F.~M.} \bibnamefont{Peeters}},
  \bibinfo{journal}{J. Appl. Phys.} \textbf{\bibinfo{volume}{118}},
  \bibinfo{pages}{154308} (\bibinfo{year}{2015}).

\bibitem[{\citenamefont{B{\o}ggild et~al.}(2017)\citenamefont{B{\o}ggild,
  Caridad, Stampfer, Calogero, Papior, and Brandbyge}}]{ncomms15783}
\bibinfo{author}{\bibfnamefont{P.}~\bibnamefont{B{\o}ggild}},
  \bibinfo{author}{\bibfnamefont{J.~M.} \bibnamefont{Caridad}},
  \bibinfo{author}{\bibfnamefont{C.}~\bibnamefont{Stampfer}},
  \bibinfo{author}{\bibfnamefont{G.}~\bibnamefont{Calogero}},
  \bibinfo{author}{\bibfnamefont{N.~R.} \bibnamefont{Papior}},
  \bibnamefont{and}
  \bibinfo{author}{\bibfnamefont{M.}~\bibnamefont{Brandbyge}},
  \bibinfo{journal}{Nature Communications} \textbf{\bibinfo{volume}{8}},
  \bibinfo{pages}{15783} (\bibinfo{year}{2017}).

\bibitem[{\citenamefont{Zhang et~al.}(2017{\natexlab{a}})\citenamefont{Zhang,
  Zhu, Yang, and Chang}}]{zhang2017}
\bibinfo{author}{\bibfnamefont{S.-H.} \bibnamefont{Zhang}},
  \bibinfo{author}{\bibfnamefont{J.-J.} \bibnamefont{Zhu}},
  \bibinfo{author}{\bibfnamefont{W.}~\bibnamefont{Yang}}, \bibnamefont{and}
  \bibinfo{author}{\bibfnamefont{K.}~\bibnamefont{Chang}}, \bibinfo{journal}{2D
  Materials} \textbf{\bibinfo{volume}{4}}, \bibinfo{pages}{035005}
  (\bibinfo{year}{2017}{\natexlab{a}}).

\bibitem[{\citenamefont{Hills et~al.}(2017)\citenamefont{Hills, Kusmartseva,
  and Kusmartsev}}]{PhysRevB.95.214103}
\bibinfo{author}{\bibfnamefont{R.~D.~Y.} \bibnamefont{Hills}},
  \bibinfo{author}{\bibfnamefont{A.}~\bibnamefont{Kusmartseva}},
  \bibnamefont{and} \bibinfo{author}{\bibfnamefont{F.~V.}
  \bibnamefont{Kusmartsev}}, \bibinfo{journal}{Phys. Rev. B}
  \textbf{\bibinfo{volume}{95}}, \bibinfo{pages}{214103}
  (\bibinfo{year}{2017}).

\bibitem[{\citenamefont{Lee et~al.}(2015)\citenamefont{Lee, Park, and
  Lee}}]{LeeNatPhys2015}
\bibinfo{author}{\bibfnamefont{G.-H.} \bibnamefont{Lee}},
  \bibinfo{author}{\bibfnamefont{G.-H.} \bibnamefont{Park}}, \bibnamefont{and}
  \bibinfo{author}{\bibfnamefont{H.-J.} \bibnamefont{Lee}},
  \bibinfo{journal}{Nat. Phys.} \textbf{\bibinfo{volume}{11}},
  \bibinfo{pages}{925} (\bibinfo{year}{2015}).

\bibitem[{\citenamefont{Chen et~al.}(2016)\citenamefont{Chen, Han, Elahi,
  Habib, Wang, Wen, Gao, Taniguchi, Watanabe, Hone et~al.}}]{ChenScience2016}
\bibinfo{author}{\bibfnamefont{S.}~\bibnamefont{Chen}},
  \bibinfo{author}{\bibfnamefont{Z.}~\bibnamefont{Han}},
  \bibinfo{author}{\bibfnamefont{M.~M.} \bibnamefont{Elahi}},
  \bibinfo{author}{\bibfnamefont{K.~M.~M.} \bibnamefont{Habib}},
  \bibinfo{author}{\bibfnamefont{L.}~\bibnamefont{Wang}},
  \bibinfo{author}{\bibfnamefont{B.}~\bibnamefont{Wen}},
  \bibinfo{author}{\bibfnamefont{Y.}~\bibnamefont{Gao}},
  \bibinfo{author}{\bibfnamefont{T.}~\bibnamefont{Taniguchi}},
  \bibinfo{author}{\bibfnamefont{K.}~\bibnamefont{Watanabe}},
  \bibinfo{author}{\bibfnamefont{J.}~\bibnamefont{Hone}}, \bibnamefont{et~al.},
  \bibinfo{journal}{Science} \textbf{\bibinfo{volume}{353}},
  \bibinfo{pages}{1522} (\bibinfo{year}{2016}).

\bibitem[{\citenamefont{Hassler et~al.}(2010)\citenamefont{Hassler, Akhmerov,
  and Beenakker}}]{HasslerPRB2010}
\bibinfo{author}{\bibfnamefont{F.}~\bibnamefont{Hassler}},
  \bibinfo{author}{\bibfnamefont{A.~R.} \bibnamefont{Akhmerov}},
  \bibnamefont{and} \bibinfo{author}{\bibfnamefont{C.~W.~J.}
  \bibnamefont{Beenakker}}, \bibinfo{journal}{Phys. Rev. B}
  \textbf{\bibinfo{volume}{82}}, \bibinfo{pages}{125423}
  (\bibinfo{year}{2010}).

\bibitem[{\citenamefont{P¨¦terfalvi et~al.}(2012)\citenamefont{P¨¦terfalvi,
  Oroszl¨¢ny, Lambert, and Cserti}}]{intraband2012}
\bibinfo{author}{\bibfnamefont{C.~G.} \bibnamefont{P¨¦terfalvi}},
  \bibinfo{author}{\bibfnamefont{L.}~\bibnamefont{Oroszl¨¢ny}},
  \bibinfo{author}{\bibfnamefont{C.~J.} \bibnamefont{Lambert}},
  \bibnamefont{and} \bibinfo{author}{\bibfnamefont{J.}~\bibnamefont{Cserti}},
  \bibinfo{journal}{New Journal of Physics} \textbf{\bibinfo{volume}{14}},
  \bibinfo{pages}{063028} (\bibinfo{year}{2012}).

\bibitem[{\citenamefont{Pereira et~al.}(2009)\citenamefont{Pereira,
  Castro~Neto, and Peres}}]{PhysRevB.80.045401}
\bibinfo{author}{\bibfnamefont{V.~M.} \bibnamefont{Pereira}},
  \bibinfo{author}{\bibfnamefont{A.~H.} \bibnamefont{Castro~Neto}},
  \bibnamefont{and} \bibinfo{author}{\bibfnamefont{N.~M.~R.}
  \bibnamefont{Peres}}, \bibinfo{journal}{Phys. Rev. B}
  \textbf{\bibinfo{volume}{80}}, \bibinfo{pages}{045401}
  (\bibinfo{year}{2009}).

\bibitem[{\citenamefont{Settnes et~al.}(2014)\citenamefont{Settnes, Power,
  Petersen, and Jauho}}]{SettnesPRL2014}
\bibinfo{author}{\bibfnamefont{M.}~\bibnamefont{Settnes}},
  \bibinfo{author}{\bibfnamefont{S.~R.} \bibnamefont{Power}},
  \bibinfo{author}{\bibfnamefont{D.~H.} \bibnamefont{Petersen}},
  \bibnamefont{and} \bibinfo{author}{\bibfnamefont{A.-P.} \bibnamefont{Jauho}},
  \bibinfo{journal}{Phys. Rev. Lett.} \textbf{\bibinfo{volume}{112}},
  \bibinfo{pages}{096801} (\bibinfo{year}{2014}).

\bibitem[{\citenamefont{Dean et~al.}(2010)\citenamefont{Dean, Young, Meric,
  Lee, Wang, Sorgenfrei, Watanabe, Taniguchi, Kim, Shepard
  et~al.}}]{DeanYoungMericEtAl2010}
\bibinfo{author}{\bibfnamefont{C.~R.} \bibnamefont{Dean}},
  \bibinfo{author}{\bibfnamefont{A.~F.} \bibnamefont{Young}},
  \bibinfo{author}{\bibfnamefont{I.}~\bibnamefont{Meric}},
  \bibinfo{author}{\bibfnamefont{C.}~\bibnamefont{Lee}},
  \bibinfo{author}{\bibfnamefont{L.}~\bibnamefont{Wang}},
  \bibinfo{author}{\bibfnamefont{S.}~\bibnamefont{Sorgenfrei}},
  \bibinfo{author}{\bibfnamefont{K.}~\bibnamefont{Watanabe}},
  \bibinfo{author}{\bibfnamefont{T.}~\bibnamefont{Taniguchi}},
  \bibinfo{author}{\bibfnamefont{P.}~\bibnamefont{Kim}},
  \bibinfo{author}{\bibfnamefont{K.~L.} \bibnamefont{Shepard}},
  \bibnamefont{et~al.}, \bibinfo{journal}{Nature Nanotechnology}
  \textbf{\bibinfo{volume}{5}}, \bibinfo{pages}{722} (\bibinfo{year}{2010}).

\bibitem[{\citenamefont{Park et~al.}(2009)\citenamefont{Park, Son, Yang, Cohen,
  and Louie}}]{PhysRevLett.103.046808}
\bibinfo{author}{\bibfnamefont{C.-H.} \bibnamefont{Park}},
  \bibinfo{author}{\bibfnamefont{Y.-W.} \bibnamefont{Son}},
  \bibinfo{author}{\bibfnamefont{L.}~\bibnamefont{Yang}},
  \bibinfo{author}{\bibfnamefont{M.~L.} \bibnamefont{Cohen}}, \bibnamefont{and}
  \bibinfo{author}{\bibfnamefont{S.~G.} \bibnamefont{Louie}},
  \bibinfo{journal}{Phys. Rev. Lett.} \textbf{\bibinfo{volume}{103}},
  \bibinfo{pages}{046808} (\bibinfo{year}{2009}).

\bibitem[{\citenamefont{Brey and Fertig}(2009)}]{PhysRevLett.103.046809}
\bibinfo{author}{\bibfnamefont{L.}~\bibnamefont{Brey}} \bibnamefont{and}
  \bibinfo{author}{\bibfnamefont{H.~A.} \bibnamefont{Fertig}},
  \bibinfo{journal}{Phys. Rev. Lett.} \textbf{\bibinfo{volume}{103}},
  \bibinfo{pages}{046809} (\bibinfo{year}{2009}).

\bibitem[{\citenamefont{Barbier et~al.}(2008)\citenamefont{Barbier, Peeters,
  Vasilopoulos, and Pereira}}]{PhysRevB.77.115446}
\bibinfo{author}{\bibfnamefont{M.}~\bibnamefont{Barbier}},
  \bibinfo{author}{\bibfnamefont{F.~M.} \bibnamefont{Peeters}},
  \bibinfo{author}{\bibfnamefont{P.}~\bibnamefont{Vasilopoulos}},
  \bibnamefont{and} \bibinfo{author}{\bibfnamefont{J.~M.}
  \bibnamefont{Pereira}}, \bibinfo{journal}{Phys. Rev. B}
  \textbf{\bibinfo{volume}{77}}, \bibinfo{pages}{115446}
  (\bibinfo{year}{2008}).

\bibitem[{\citenamefont{Barbier et~al.}(2009)\citenamefont{Barbier,
  Vasilopoulos, Peeters, and Pereira}}]{PhysRevB.79.155402}
\bibinfo{author}{\bibfnamefont{M.}~\bibnamefont{Barbier}},
  \bibinfo{author}{\bibfnamefont{P.}~\bibnamefont{Vasilopoulos}},
  \bibinfo{author}{\bibfnamefont{F.~M.} \bibnamefont{Peeters}},
  \bibnamefont{and} \bibinfo{author}{\bibfnamefont{J.~M.}
  \bibnamefont{Pereira}}, \bibinfo{journal}{Phys. Rev. B}
  \textbf{\bibinfo{volume}{79}}, \bibinfo{pages}{155402}
  (\bibinfo{year}{2009}).

\bibitem[{\citenamefont{Rold¨¢n et~al.}(2015)\citenamefont{Rold¨¢n,
  Castellanos-Gomez, Cappelluti, and Guinea}}]{RafaelJPCM2015}
\bibinfo{author}{\bibfnamefont{R.}~\bibnamefont{Rold¨¢n}},
  \bibinfo{author}{\bibfnamefont{A.}~\bibnamefont{Castellanos-Gomez}},
  \bibinfo{author}{\bibfnamefont{E.}~\bibnamefont{Cappelluti}},
  \bibnamefont{and} \bibinfo{author}{\bibfnamefont{F.}~\bibnamefont{Guinea}},
  \bibinfo{journal}{Journal of Physics: Condensed Matter}
  \textbf{\bibinfo{volume}{27}}, \bibinfo{pages}{313201}
  (\bibinfo{year}{2015}).

\bibitem[{\citenamefont{Zhang et~al.}(2017{\natexlab{b}})\citenamefont{Zhang,
  Yang, and Chang}}]{ZhangPRB2017}
\bibinfo{author}{\bibfnamefont{S.-H.} \bibnamefont{Zhang}},
  \bibinfo{author}{\bibfnamefont{W.}~\bibnamefont{Yang}}, \bibnamefont{and}
  \bibinfo{author}{\bibfnamefont{K.}~\bibnamefont{Chang}},
  \bibinfo{journal}{Phys. Rev. B} \textbf{\bibinfo{volume}{95}},
  \bibinfo{pages}{075421} (\bibinfo{year}{2017}{\natexlab{b}}).

\bibitem[{\citenamefont{{Forsythe} et~al.}(2017)\citenamefont{{Forsythe},
  {Zhou}, {Taniguchi}, {Watanabe}, {Pasupathy}, {Moon}, {Koshino}, {Kim}, and
  {Dean}}}]{171001365F}
\bibinfo{author}{\bibfnamefont{C.}~\bibnamefont{{Forsythe}}},
  \bibinfo{author}{\bibfnamefont{X.}~\bibnamefont{{Zhou}}},
  \bibinfo{author}{\bibfnamefont{T.}~\bibnamefont{{Taniguchi}}},
  \bibinfo{author}{\bibfnamefont{K.}~\bibnamefont{{Watanabe}}},
  \bibinfo{author}{\bibfnamefont{A.}~\bibnamefont{{Pasupathy}}},
  \bibinfo{author}{\bibfnamefont{P.}~\bibnamefont{{Moon}}},
  \bibinfo{author}{\bibfnamefont{M.}~\bibnamefont{{Koshino}}},
  \bibinfo{author}{\bibfnamefont{P.}~\bibnamefont{{Kim}}}, \bibnamefont{and}
  \bibinfo{author}{\bibfnamefont{C.~R.} \bibnamefont{{Dean}}},
  \bibinfo{journal}{ArXiv}  (\bibinfo{year}{2017}), \eprint{1710.01365}.

\bibitem[{\citenamefont{Yuan et~al.}(2016)\citenamefont{Yuan, Zhang, Liu,
  Narayan, Song, Shen, Sui, Xu, Yu, An et~al.}}]{YuanNPG2016}
\bibinfo{author}{\bibfnamefont{X.}~\bibnamefont{Yuan}},
  \bibinfo{author}{\bibfnamefont{C.}~\bibnamefont{Zhang}},
  \bibinfo{author}{\bibfnamefont{Y.}~\bibnamefont{Liu}},
  \bibinfo{author}{\bibfnamefont{A.}~\bibnamefont{Narayan}},
  \bibinfo{author}{\bibfnamefont{C.}~\bibnamefont{Song}},
  \bibinfo{author}{\bibfnamefont{S.}~\bibnamefont{Shen}},
  \bibinfo{author}{\bibfnamefont{X.}~\bibnamefont{Sui}},
  \bibinfo{author}{\bibfnamefont{J.}~\bibnamefont{Xu}},
  \bibinfo{author}{\bibfnamefont{H.}~\bibnamefont{Yu}},
  \bibinfo{author}{\bibfnamefont{Z.}~\bibnamefont{An}}, \bibnamefont{et~al.},
  \bibinfo{journal}{Npg Asia Materials} \textbf{\bibinfo{volume}{8}},
  \bibinfo{pages}{e325} (\bibinfo{year}{2016}).

\bibitem[{\citenamefont{Qiu et~al.}(2016)\citenamefont{Qiu, Du, Charnas, Zhou,
  Jin, Luo, Zemlyanov, Xu, Cheng, and Ye}}]{NL6b02629}
\bibinfo{author}{\bibfnamefont{G.}~\bibnamefont{Qiu}},
  \bibinfo{author}{\bibfnamefont{Y.}~\bibnamefont{Du}},
  \bibinfo{author}{\bibfnamefont{A.}~\bibnamefont{Charnas}},
  \bibinfo{author}{\bibfnamefont{H.}~\bibnamefont{Zhou}},
  \bibinfo{author}{\bibfnamefont{S.}~\bibnamefont{Jin}},
  \bibinfo{author}{\bibfnamefont{Z.}~\bibnamefont{Luo}},
  \bibinfo{author}{\bibfnamefont{D.~Y.} \bibnamefont{Zemlyanov}},
  \bibinfo{author}{\bibfnamefont{X.}~\bibnamefont{Xu}},
  \bibinfo{author}{\bibfnamefont{G.~J.} \bibnamefont{Cheng}}, \bibnamefont{and}
  \bibinfo{author}{\bibfnamefont{P.~D.} \bibnamefont{Ye}},
  \bibinfo{journal}{Nano Letters} \textbf{\bibinfo{volume}{16}},
  \bibinfo{pages}{7364} (\bibinfo{year}{2016}).

\bibitem[{\citenamefont{Neupane et~al.}(2014)\citenamefont{Neupane, Xu, Sankar,
  Alidoust, Bian, Liu, Belopolski, Chang, Jeng, Lin et~al.}}]{NeupaneNC2014}
\bibinfo{author}{\bibfnamefont{M.}~\bibnamefont{Neupane}},
  \bibinfo{author}{\bibfnamefont{S.-Y.} \bibnamefont{Xu}},
  \bibinfo{author}{\bibfnamefont{R.}~\bibnamefont{Sankar}},
  \bibinfo{author}{\bibfnamefont{N.}~\bibnamefont{Alidoust}},
  \bibinfo{author}{\bibfnamefont{G.}~\bibnamefont{Bian}},
  \bibinfo{author}{\bibfnamefont{C.}~\bibnamefont{Liu}},
  \bibinfo{author}{\bibfnamefont{I.}~\bibnamefont{Belopolski}},
  \bibinfo{author}{\bibfnamefont{T.-R.} \bibnamefont{Chang}},
  \bibinfo{author}{\bibfnamefont{H.-T.} \bibnamefont{Jeng}},
  \bibinfo{author}{\bibfnamefont{H.}~\bibnamefont{Lin}}, \bibnamefont{et~al.},
  \bibinfo{journal}{Nature Communications} \textbf{\bibinfo{volume}{5}},
  \bibinfo{pages}{3786} (\bibinfo{year}{2014}).

\bibitem[{\citenamefont{Leder et~al.}(2014)\citenamefont{Leder, Grossert, and
  Weitz}}]{LederNC2014}
\bibinfo{author}{\bibfnamefont{M.}~\bibnamefont{Leder}},
  \bibinfo{author}{\bibfnamefont{C.}~\bibnamefont{Grossert}}, \bibnamefont{and}
  \bibinfo{author}{\bibfnamefont{M.}~\bibnamefont{Weitz}},
  \bibinfo{journal}{Nature Communications} \textbf{\bibinfo{volume}{5}},
  \bibinfo{pages}{3327} (\bibinfo{year}{2014}).

\bibitem[{\citenamefont{Tarruell et~al.}(2012)\citenamefont{Tarruell, Greif,
  Uehlinger, Jotzu, and Esslinger}}]{TarruellN2012}
\bibinfo{author}{\bibfnamefont{L.}~\bibnamefont{Tarruell}},
  \bibinfo{author}{\bibfnamefont{D.}~\bibnamefont{Greif}},
  \bibinfo{author}{\bibfnamefont{T.}~\bibnamefont{Uehlinger}},
  \bibinfo{author}{\bibfnamefont{G.}~\bibnamefont{Jotzu}}, \bibnamefont{and}
  \bibinfo{author}{\bibfnamefont{T.}~\bibnamefont{Esslinger}},
  \bibinfo{journal}{Nature} \textbf{\bibinfo{volume}{483}},
  \bibinfo{pages}{302} (\bibinfo{year}{2012}).

\bibitem[{\citenamefont{Katsnelson et~al.}(2006)\citenamefont{Katsnelson,
  Novoselov, and Geim}}]{KatsnelsonNatPhys2006}
\bibinfo{author}{\bibfnamefont{M.~I.} \bibnamefont{Katsnelson}},
  \bibinfo{author}{\bibfnamefont{K.~S.} \bibnamefont{Novoselov}},
  \bibnamefont{and} \bibinfo{author}{\bibfnamefont{A.~K.} \bibnamefont{Geim}},
  \bibinfo{journal}{Nat. Phys.} \textbf{\bibinfo{volume}{2}},
  \bibinfo{pages}{620} (\bibinfo{year}{2006}).

\bibitem[{\citenamefont{Cheianov and Fal'ko}(2006)}]{CheianovPRB2006}
\bibinfo{author}{\bibfnamefont{V.~V.} \bibnamefont{Cheianov}} \bibnamefont{and}
  \bibinfo{author}{\bibfnamefont{V.~I.} \bibnamefont{Fal'ko}},
  \bibinfo{journal}{Phys. Rev. B} \textbf{\bibinfo{volume}{74}},
  \bibinfo{pages}{041403} (\bibinfo{year}{2006}).

\end{thebibliography}

\end{document}